\documentstyle[prl,aps,psfig]{revtex}
\begin{document}
\draft
\title{The Non-Local Nature of the Nuclear Force and Its Impact
on Nuclear Structure}
\author{R. Machleidt~\footnote{e-mail address: machleid@phys.uidaho.edu},
F. Sammarruca, and Y. Song}
\address{Department of Physics, University of Idaho, Moscow,
ID 83843, U. S. A.}

\date{\today}

\maketitle

\begin{abstract}
We calculate the triton binding energy with a non-local NN
potential that fits the world NN data below 350 MeV
with the almost perfect $\chi^2$/datum of 1.03.
The non-locality is derived from relativistic meson
field theory.
The result obtained in a 34-channel, charge-dependent
Faddeev calculation is 8.00 MeV, which is 0.4 MeV
above the predictions by local NN potentials.
The increase in binding energy can be clearly
attributed to the off-shell behavior of the non-local
potential.
Our result cuts in half the discrepancy between theory
and experiment established from local NN potentials.
Implications for other areas of microscopic nuclear structure,
in which underbinding is a traditional problem, are discussed.
\end{abstract}


\twocolumn

One of the most fundamental traditional goals of theoretical nuclear physics
is to explain the properties of atomic nuclei in terms of the
elementary interactions between nucleons.
For this program, the basic interaction between {\it two} nucleons
is the most important ingredient.
But even if one considers only two-nucleon interactions,
the nuclear many-body problem does not have a unique
solution.
This non-uniqueness is due to the fact that, in the many-body
system, the nucleon-nucleon (NN) interaction contributes
also ``off the energy shell'' (off-shell).

By construction, NN interactions reproduce the two-nucleon
scattering data and the properties of the deuteron.
Assuming the existence of NN scattering data of increasing
quantity and quality, the NN interaction can be fixed with arbitrary
accuracy---``on the energy shell'' (on-shell); i.~e.,
for processes in which the two nucleons have the same energy
before and after the interaction, like in free-space
NN scattering.

In a nucleus with more than two nucleons (A $>2$),
the energy of the A-particle system is conserved.
However, that does not imply that energy is conserved
in any individual
interaction between two nucleons in the nucleus.
Thus, in a many-body system,
two nucleons may have different energies before
and after they interact; i.~e., their mutual
interaction may be off the energy shell.
Therefore, the calculation of, e.~g., the binding energy of
an A-particle nucleus involves the off-shell NN interaction,
which is  empirically undetermined; only theory can
provide it.

The off-shell problem in microscopic nuclear structure
has been known for several decades~\cite{dSF74}.
However, in spite of many efforts,
it has not been possible, to date, to precisely pin down
the off-shell effect on, e.~g., the binding energy
of a nucleus.
Past work on this topic has suffered from
two major drawbacks.
In some work~\cite{HT70}, the NN interactions used
were realistic, but not exactly identical on-shell (or not exactly
``phase-equivalent'', where phase refers to phase shifts
of free-space NN scattering).
In other works~\cite{Coe70,HT71},
phase-equivalent potentials have been constructed
by some mathematical methods,
but it is doubtful whether the constructed off-shell behavior
is realistic, i.~e., resembles anything that would be created
by mechanism underlying the nuclear force.

The problems in past studies may give us some idea of which
minimal requirements should be met by a reliable investigation
of the issue:
The NN potentials considered should predict the NN observables
identically {\it and} in accurate agreement with the data.
Furthermore, the potentials should have some basis in theory.

Recent substantial progress in the field of nuclear few-body physics
has finally set the stage for an investigation
of off-shell effects in microscopic nuclear structure
which can fulfil the above requirements.
In 1993, the Nijmegen group has published a phase-shift analysis
of all proton-proton and neutron-proton data below
350 MeV lab.\ energy with a $\chi^2$ per datum of 0.99
for 4301 data.
Based upon these data, charge-dependent NN potentials have
been constructed by the Nijmegen~\cite{Sto94} and the
Argonne~\cite{WSS95} groups which reproduce the NN data with
a $\chi^2$/datum of 1.03 and 1.09, respectively.
This agreement between the potential predictions and the
data as well as the agreement among the various potentials
is, on statistical grounds, as accurate as it can be.

An appropriate sample nucleus for microscopic test calculations
is the triton ($^3$H). It is the smallest A $>2$ nucleus
which does not involve the Coulomb force, and rigorous,
charge-dependent calculations of this nucleus are now-a-days
a routine matter.
At first glance, the triton may appear too simplistic
to represent a reliable test ground for some general features of microscopic
nuclear structure; however, Gl\"ockle and Kamada~\cite{GK93}
have shown that the rigorous solutions
of larger few-nucleon problems (which are very involved and expensive)
have essentially the same characteristics as the three-nucleon
system.
Moreover, there are even obvious parallels between results
for the triton, on the one hand,
and predictions for  nuclear matter and excited
nuclei,
on the other~\cite{foot1}.

Friar {\it et al.}~\cite{Fri93} have calculated the binding
energy of the triton (in charge-dependent 34-channel
Faddeev calculations)
applying the new, high-quality potentials and obtained
almost identical results for the various local models,
namely,
$7.62\pm 0.01$ MeV (experimental value: 8.48 MeV),
where the uncertainty of $\pm 0.01$ MeV
is the variation of the predictions which occurs when
different local potentials are used.
The smallness of the variation is due to the fact that all the local
potentials used in the study
have essentially the same off-shell behavior.

All new high-quality NN potentials
use the local version of the one-pion-exchange potential
for the long-range part of the interaction which is,
e.~g., for $pp$ scattering:
\begin{eqnarray}
V_\pi^{(loc)}({\bf r}) = &
\frac{g^2_\pi}{12\pi}
\left(\frac{m_\pi}{2M}\right)^2
\left[
\left(
\frac{e^{-m_\pi r}}{r}
-\frac{4\pi}{m_\pi^2}\delta^{(3)}({\bf r})
\right)
\mbox{\boldmath $\sigma_{1} \cdot \sigma_{2}$}
\right.
\nonumber \\
& +
\left.
\left(1+\frac{3}{m_\pi r}+\frac{3}{(m_\pi r)^2}\right)
\frac{e^{-m_\pi r}}{r}
\mbox{\boldmath $S_{12}$}
\right] \; ,
\end{eqnarray}
where $m_\pi$
denotes the neutral pion mass and $M$ the proton mass.
For $np$ scattering the appropriate combination of neutral
and charged pion exchange is used which creates the charge-dependence
in the models.
The intermediate and short-range parts
are parametrized in different ways.
Here, the Argonne $V_{18}$ potential~\cite{WSS95} uses local functions
of Woods-Saxon type,
while the other potentials apply local Yukawas of either multiples
of the pion mass (Reid'93~\cite{Sto94}) or of the empirical
masses of existing mesons and meson distributions
(Nijm-II~\cite{Sto94}).
All potentials are regularized at short distances by either
exponential ($V_{18}$, Nijm-II) or dipole (Reid'93)
form factors (which are also local).

Ever since NN potentials have been developed,
local potentials have enjoyed great popularity
because they are easy to apply in configuration-space
calculations.
Note, however, that numerical ease is not a proof for the local nature
of the nuclear force.
In fact, any deeper insight into the reaction mechanisms underlying
the nuclear force suggests a non-local character.
In particular, the composite structure of hadrons should lead
to large non-localities at short range~\cite{VS95}.
But, even the conventional and well-established
meson theory of nuclear forces---when derived properly and without
crude approximations---creates a non-local interaction.
In this paper, we will focus on this ``simplest''
source of non-locality.

Common Lagrangians for meson-nucleon coupling are:
\begin{eqnarray}
{\cal L}_{ps}&=& -g_{ps}\bar{\psi}
i\gamma^{5}\psi\varphi^{(ps)}
\\
{\cal L}_{s}&=& g_{s}\bar{\psi}\psi\varphi^{(s)}
\\
{\cal L}_{v}&=&g_{v}\bar{\psi}\gamma^{\mu}\psi\varphi^{(v)}_{\mu}
+\frac{f_{v}}{4M} \bar{\psi}\sigma^{\mu\nu}\psi(\partial_{\mu}
\varphi_{\nu}^{(v)}
-\partial_{\nu}\varphi_{\mu}^{(v)})
\end{eqnarray}
where
$ps$, $s$, and $v$ denote pseudoscalar, scalar, and vector
couplings/fields, respectively.

The lowest order contributions to the nuclear force from the above
Lagrangians are the second-order Feynman diagrams
which, in the c.~m. system of the two interacting nucleons, produce
the amplitude:
\begin{equation}
{\cal A}_{\alpha}(q',q) =
\frac{\bar{u}_1({\bf q'})\Gamma_1^{(\alpha)} u_1({\bf q}) P_\alpha
\bar{u}_2(-{\bf q'})\Gamma_2^{(\alpha)} u_2(-{\bf q})}
{(q'-q)^2-m_\alpha^2} \; ,
\end{equation}
where $\Gamma_i^{(\alpha)}$ ($i=1,2$) are vertices
derived from the above Lagrangians, $u_i$ Dirac spinors representing
the interacting nucleons, and $q$ and $q'$ their relative momenta
in the initial and final states, respectively; $P_\alpha$
divided by the denominator is  the
meson propagator.

The simplest meson-exchange model for the nuclear force is
the one-boson-exchange (OBE) potential~\cite{Mac89,Mac93} which sums over
several second-order diagrams, each representing
the single exchange of a different boson, $\alpha$:
\begin{equation}
V({\bf q'},{\bf q}) =
\sqrt{\frac{M}{E'}}
\sqrt{\frac{M}{E}}
\sum_\alpha i{\cal A}_{\alpha}({\bf q'},{\bf q})
F_{\alpha}^2({\bf q'},{\bf q}) \; .
\end{equation}
As customary, we included form factors,
$F_{\alpha}({\bf q'},{\bf q})$, applied to the
meson-nucleon vertices, and a square-root factor
$M/
\sqrt{E'E}$
(with $E=\sqrt{M^2+{\bf q}^2}$
and $E'=\sqrt{M^2+{\bf q'}^2}$).
The form factors regularize the amplitudes for large momenta
(short distances) and account for the extended structure of
nucleons in a phenomenological way.
The square root factors make it possible to cast the
unitarizing,
relativistic, three-dimensional Blankenbecler-Sugar equation
for the scattering amplitude
(a reduced version of the four-dimensional Bethe-Salpeter
equation)
into a form
which is identical to the (non-relativistic) Lippmann-Schwinger
equation~\cite{Mac89,Mac93}. Thus, Eq.~(6) defines a relativistic
potential which can be consistently applied in conventional, non-relativistic
nuclear structure.

Clearly, the Feynman amplitudes, Eq.~(5), are in general non-local
expressions; i.~e., Fourier transform into
configuration space
will yield functions of $r$ and $r'$, the relative
distances between the two in- and out-going nucleons,
respectively.
The square root factors create additional non-locality,
as pointed out by Gl\"ockle and Kamada~\cite{GK94}.

\begin{figure}[h]
\vspace{1.2cm}\hspace*{1cm}\psfig{figure=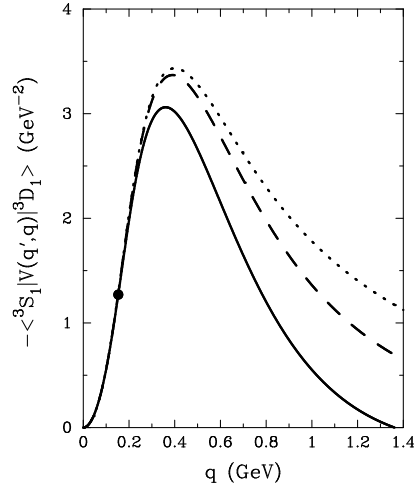,height=5.5cm}
\caption{Half off-shell $^3S_1$--$^3D_1$ amplitude
for the relativistic CD-Bonn potential (solid line), Eq.~(6).
The dashed (dotted) curve is obtained when the local
approximation, Eq.~(8), is used for OPE (OPE and one-$\rho$
exchange). $q'=153$ MeV.}
\end{figure}

While for heavy vector-meson exchange (corresponding to short distances)
non-locality appears quite plausible, we have to stress here
that even the one-pion-exchange (OPE) Feynman amplitude is non-local.
This is important because the pion creates the dominant part of the tensor
force
which plays a crucial role in nuclear structure.

Applying $\Gamma^{(\pi)}=g_\pi \gamma_5$ in Eq.~(5),
yields the Feynman amplitude for neutral pion exchange in $pp$ scattering,
\begin{eqnarray}
i{\cal A}_\pi ({\bf q'}, {\bf q})  = & -\frac{g^2_\pi}{4M^2}
\frac{(E'+M)(E+M)}
{({\bf q'}-{\bf q})^2+m_\pi^2}
\left(
 \frac{\mbox{\boldmath $\sigma_{1} \cdot $} {\bf q'}}{E'+M}
-
 \frac{\mbox{\boldmath $\sigma_{1} \cdot $} {\bf q}}{E+M}
\right)
\nonumber \\
 &
\times
\left(
 \frac{\mbox{\boldmath $\sigma_{2} \cdot $} {\bf q'}}{E'+M}
-
 \frac{\mbox{\boldmath $\sigma_{2} \cdot $} {\bf q}}{E+M}
\right) \; .
\end{eqnarray}
In static approximation, i.~e., for $E'\approx E \approx M$, this reduces to
\begin{equation}
V_\pi^{(loc)}({\bf k})  =  -\frac{g_{\pi}^{2}}{4M^{2}}
 \frac{{\mbox {\boldmath $(\sigma_{1} \cdot $}} {\bf k)}
{\mbox{\boldmath $(\sigma_{2} \cdot $}} {\bf k)}}
 {{\bf k}^{2}+m_{\pi}^{2}}
\end{equation}
with ${\bf k} = {\bf q'} - {\bf q}$; this is nothing but the
Fourier transform of the local OPE potential, $V_\pi^{(loc)}({\bf r})$,
given in Eq.~(1).
Notice also that on-shell, i.~e., for $|{\bf q'}|=|{\bf q}|$,
$V^{(loc)}_\pi$ equals $i{\cal A}_\pi$.
Thus, the non-locality affects the OPE potential
only off-shell.

In Fig.~1, we show the half off-shell
$^3S_1$--$^3D_1$
potential that can be produced only by tensor forces.
The on-shell momentum $q'$ is held fixed at 153 MeV
(equivalent to 50 MeV lab.\ energy),
while the off-shell momentum $q$ runs from zero
to 1400 MeV.
The on-shell point ($q=153$ MeV) is marked by a solid dot.
The solid curve is a relativistic OBE potential
(CD-Bonn, s.\ below), Eq.~(6). When the relativistic
OPE amplitude, Eq.~(7), is replaced by the static/local
approximation, Eq.~(8),
the dashed curve is obtained.
When this approximation is also used for the one-$\rho$
exchange, the dotted curve results.
It is clearly seen that the static/local approximation
substantially increases the tensor force off-shell.

 From the discussion here, it is evident that relativity and non-locality
are intimately interwoven.
At this advanced stage of nuclear few-body physics, there is a need for
relativistic potentials, also, for reasons other than
non-locality~\cite{Fri94}. Potentials based upon
the invariant Feynman amplitudes, Eq.~(5), are examples for
relativistic potentials.
A very quantitative model of this kind will be given below.

We believe that the non-localities created by
relativistic meson exchange are ``real'' and deserve attention.
An important question is:
What is their impact on microscopic
nuclear structure calculations?

To investigate this point, we have constructed a new
relativistic OBE potential
based upon Eqs.~(5) and (6) of very high precision.
The potential (dubbed ``CD-Bonn'')
is charge-dependent due to nucleon and pion
mass splitting; therefore, a $pp$,
$np$, and $nn$ potential are provided.

To meet the requirements for a reliable investigation pointed
out above, we have fitted the 4301 $pp$ and $np$ data below
350 MeV lab.\ energy with a $\chi^2$/datum of 1.03~\cite{foot2}
(i.~e., with the same accuracy as the new local high-quality
potentials~\cite{Sto94,WSS95}).
As pointed out in Refs.~\cite{Sto93,Sto94}, this high accuracy
cannot be achieved with the usual, about a dozen, parameters of the
conventional OBE model. Some additional fit freedom is needed,
for which we choose to adjust
the fictitious $\sigma$ boson
individually in each partial wave.
Physical justification for this procedure
comes from the fact that---based upon
the more realistic meson model for the nuclear force
which includes
all important multi-meson exchanges~\cite{MHE87}---the
one-$\sigma$ exchange in the OBE model stands
for the sum
of all higher order diagrams, and not
just for the
$2\pi$-exchange (as commonly believed).
Of course, this is
\begin{table}[h]
\caption{Recent high-precision NN potentials and predictions for
the two- and three-nucleon system}
\begin{tabular}{lcccc}
              & CD-Bonn  &   Nijm-II  &   Reid'93 & $V_{18}$ \\
\hline
\hline
Character  & non-local   & local           & local          & local    \\
$\chi^2$/datum  & 1.03 & 1.03     & 1.03           & 1.09     \\
$g^2_\pi/4\pi$  & 13.6 & 13.6     & 13.6           & 13.6     \\
\hline
Deuteron properties: \\
Quadr.\ moment (fm$^2$) & 0.270 & 0.271       & 0.270          & 0.270    \\
Asymptotic D/S state & 0.0255  & 0.0252          & 0.0251         & 0.0250   \\
D-state probab.\ (\%) & 4.83& 5.64            & 5.70           & 5.76     \\
\hline
Triton binding (MeV): \\
non-rel.\ calculation & 8.00   & 7.62            & 7.63           & 7.62     \\
relativ.\ calculation & 8.19 & -- & -- & -- \\
\end{tabular}
\end{table}
\noindent
a very crude approximation and, therefore,
typical
discrepancies occur in various partial waves
(cf.\ Fig.~11 of
Ref.~\cite{MHE87}),
which can be removed by
individual adjustments of the $\sigma$ boson.
More details concerning the new
CD-Bonn potential will be published
elsewhere~\cite{Mac96}.

In Table I (upper part), we summarize two-nucleon properties predicted
by the new CD-Bonn potential and compare with
the other recent high-quality potentials.
Using the same $\pi NN$ coupling constant,
all potentials predict almost identical deuteron observables
(quadrupole moment
and asymptotic D/S state normalization). Note, however,
that the (un-observable) deuteron
D-state probability comes out significantly larger
for the local potentials ($\approx 5.7$\%)
as compared to
the non-local CD-Bonn potential (4.8\%).
Obviously, the deuteron D-state probability
is kind of a numerical measure for
the off-shell strength of the tensor force,
shown graphically in Fig.~1.

We have performed a (34-channel, charge-dependent)
Faddeev calculation for the triton with the new CD-Bonn
potential and obtained 8.00 MeV binding energy (cf.\ Table I).
This is 0.38 MeV more than local potentials
predict.
The unaquainted observer may be tempted to believe
that this difference of 0.38 MeV is quite small,
almost negligible. However, this is not true.
The difference between the predictions by local potentials
(7.62 MeV) and experiment (8.48 MeV)
is 0.86 MeV.
Thus, the problem with the triton binding is that
0.86 MeV cannot be explained in the simplest way,
that is all.
Therefore, any non-trivial contribution
must be measured against the 0.86 MeV gap between experiment
and simplest theory.
On this scale, the non-locality considered in this investigation
explains 44\% of the gap; i.~e., it is substantial
with respect to the remaining discrepancy.

We note that there is one new Nijmegen potential (Nijm-I~\cite{Sto94})
which has a non-local central force (but is local otherwise)
and predicts 7.72 MeV for the triton binding~\cite{Fri93},
0.10 MeV above the local benchmark.
Since in our model, all components of the nuclear force
(central, tensor, etc.) are non-local, this effect is included
in our calculations; and one may conclude that the non-locality
in the tensor force increases the binding by about 0.3 MeV.

The above three-body results were obtained by using the
conventional non-relativistic Faddeev equations.
However, since CD-Bonn is a relativistic potential,
one can also perform
a relativistic Faddeev calculation by extending
the relativistic three-dimensional Blanckenbecler-Sugar
formalism to the three-body system~\cite{SXM92}.
The binding energy prediction by CD-Bonn
then goes up to 8.19 MeV.
This further increase can be understood as an additional off-shell effect
from the relativistic two-nucleon $t$-matrix applied in the three-nucleon
system (see Ref.~\cite{SXM92} for details).

We stress that our present calculations take only the most
``primitive'' source of non-locality into account. Since meson exchange
is mainly responsible for the long and intermediate range
of the nuclear force, we do not expect it to be
the main source for non-locality. The short-range part
of the nuclear force, where the composite structure of hadrons
should play an important role, may provide much larger
non-localities.
It is an challenging topic for future research
to derive
this additional non-locality~\cite{VS95}, and test its impact on
nuclear structure predictions.

Obviously,
our results leave little room for contributions
from three-nucleon forces (3NF).
Still, this does not mean that they do not exist
in nature. In fact, the meson theory of $\pi N$ and
and NN scattering (including meson resonances and
isobar degrees of freedoms) implies a large
variety of 3NF.
However, consistent calculations which
treat the 2N and 3N system on an equal footing have shown
that large cancellations can occur between ``genuine'' 3NF contributions
and medium effects on the 2N force when inserted into the three-nucleon
system~\cite{Wei92,Pic92}.
If the 3NF is weak, then because of cancellations of this kind.
If it should turn out that these cancellations are almost perfect, then
the challenging question will be if this is just an accident
of nature (which is hard to believe) or if
we are still missing some symmetries underlying
nuclear structure.

In summary,
a non-local NN potential based upon relativistic meson
theory predicts 0.4 MeV more triton binding energy
than local NN potentials.
This result cuts in half the discrepancy
between theory and experiment established from local potentials.
Based upon nuclear matter results
and earlier calculations in finite nuclei
\cite{Jia92},
one may expect that
this new high-precision non-local potential
could also improve predictions
in other areas of microscopic nuclear struture where underbinding
is a traditional problem.

This work has been supported in part by the National Science Foundation
under Grant No.~PHY-9211607.

\end{document}